\documentclass[12pt,nofootinbib]{revtex4-2}

\usepackage{graphicx, amsmath, amssymb}

\begin{document}
\title{Staggered bosons}
\author{David Berenstein}
\address{Dept. of Physics, University of California, Santa Barbara, CA 93106}

\begin{abstract}
A model with a half boson degree of freedom per lattice site in one dimension is developed. 
The boson is protected from developing a gap by translation symmetry: while the left movers are at zero quasi-momentum,  the associated right movers are at the midpoint of  the quasi-momentum period.
The model has different properties depending on if a periodic lattice has an even or an odd number of sites and similar features are found for open boundary conditions.  A special case of the non-linear half boson model where even and odd lattice sites contribute differently to the Hamiltonian gives rise to the Toda chain and a more symmetric generalization of the Toda chain is found. Upon periodic identifications of the half bosons degrees of freedom under a shift,   the total Hilbert space has a finite dimension and can be encoded in finitely many qubits per unit length. This way one finds interesting critical spin chains, examples of which include the critical Ising model in a transverse magnetic field and the 3-state Potts model at criticality. Extensions to higher dimensions are considered. Models obtained this way automatically produce dynamical systems of gapless fractons. 
\end{abstract}

\maketitle

\section{Introduction}

Euclidean lattice gauge theories a la Wilson \cite{Wilson:1974sk}, based on a path integral formulation of quantum field theories,  have been extremely successful in solving various problems in quantum field theory that are relevant for the theory of the strong interactions (see for example \cite{Aoki:2016frl,FlavourLatticeAveragingGroup:2019iem}). 
There are many situations where a determination of answers to important physical questions with such setups is essentially impossible. These situations are present in most examples for the real time evolution of physical systems. In a path integral formulation of real time evolution, the action is imaginary and the contributions between different configurations cancel each other out. 
This is called the sign problem. These effects are physical: in quantum mechanics amplitudes can cancel each other out and this can be observed directly in interference experiments. These  arise because the standard Hilbert space formulation of quantum mechanics is based on the complex numbers. Different states can be superposed with each other with arbitrary phases to give rise to such phenomena in practice. 

On a quantum computer, at least in principle, these sign problems are solved. The quantum memory of such a computer is a Hilbert space itself, so it already encodes the possibility of superposition between different states. 
Recent advances on quantum control have made the idea of practical  quantum computers feasible for some simulation tasks in the near future. They are currently constrained in that they should not involve too many degrees of freedom (qubits). 
Computations on quantum computers are based on Hamiltonian methods rather than path integral methods. One manipulates the states in the quantum memory by {\em solving} the Schr\"odinger equation
\begin{equation}
i \partial_t | \psi\rangle = \hat H(t) | \psi\rangle
\end{equation}
from some initial state $|\psi_0\rangle$. 
The computation is performed in the manipulation of the hamiltonian $H(t)$ \footnote{In practice one approximates a desired physical setup with the limitations placed on the instantaneous form of $H$ that is given by the quantum platform one is encoding the problem in.
More precisely, evolutions for small times with a particular instantaneous $H$ give rise to quantum gates (discrete time step operations) that are combined to approximate a desired output.}.
It is the advent of quantum computers in the near future that has renewed the interest in Hamiltonian methods in quantum field theory applied to particle physics questions. 

With a view towards a  real time quantum computation in the near future, low dimensional field theories on a lattice should be studied first.  Qubit degrees of freedom are more akin to bosons rather than fermions. After all, qubit operations at different locations commute with each other rather 
than anticommute. Is it possible to simulate relativistic field theories of bosons in such a situation? The usual problem is that bosons naturally develop a gap: the mass term for the bosons is relevant and is usually generated  by interactions with other fields. 
This is usually solved by fine-tuning the lattice Hamiltonian and working in the limit where the excitations above the ground state have low energy relative to the natural scale set by the lattice.
One should also worry that bosons admit arbitrary occupation numbers and naively require an infinite Hilbert space locally. This is actually more of an issue about the infrared physics than  the UV scale (the lattice cutoff). Local large occupation numbers should cost a lot of energy, so a truncation in local energy should be doable with a finite local Hilbert space at a site. This does not forbid the collective low energy excitations from having large occupation numbers.

The purpose of this paper is to identify a new class of gapless boson theories that arise from imitating a trick that is used to study fermions on a lattice and to avoid doublers. 
This is the idea of staggered fermions \cite{Kogut:1974ag}, where different pieces of the degrees of freedom of a fermion reside at different lattice sites. This is why the bosonic degrees of freedom described here will be 
called staggered bosons. 

The main idea to consider is that in a Hamiltonian formulation of a single bosonic degree of freedom one usually has both a coordinate $x$ and its canonical conjugate $p$. From the point if view of phase space $x,p$ are just different coordinates and there is a canonical transformation $x\to p, p\to -x$ that 
can turn them into each other. If we are to split these degrees of freedom between different sites, there should be a single bosonic $q$ variable at each site, rather than two and the Poisson bracket commutation relations between nearest neighbors should differ from zero. 
Basically, in this paper a model of a boson with a half degree of freedom per lattice site is proposed. In one dimension the model gives rise to gapless degrees of freedom that are automatically protected from developing a mass. 

If we consider systems with a half fermion degree of freedom per lattice site one is usually in a system of Majorana fermions. Majorana fermions enjoy rich topological features, as shown by Kitaev \cite{Kitaev:2000nmw}. 
I will show that staggered bosons also have interesting topological features that are not shared by regular bosons and which share some similarities with Majorana fermions. 
Some interesting systems like the Toda chain can  also be expressed easily in terms of half boson degrees of freedom. Another fact I will show is that upon gauging discrete translation symmetries of these bosons, for the simplest Hamiltonians one finds critical spin chains. 
Moreover, when one pushes these ideas to higher dimensions, one naturally lands in gapless field theories associated to fractons. 

These facts makes the staggered boson idea an interesting source of non-trivial examples of interacting field theories with non trivial critical behavior.

  The paper is organized as follows. In section \ref{sec:chb} the chiral boson in one dimension is studied as an example of a system with half a bosonic degree of freedom to motivate the lattice Hamiltonians that follow afterwards. Next, in section \ref{sec:model}, the basic model of the half boson is proposed and the dispersion relation is studied. In section \ref{sec:top} topological features of the model are studied. These include the protected nature of zero modes, the number of zero modes in finite lattices (open and periodic) and the dependence of these on the number of sites.  Also the parity symmetry and construction of the model as projections from regular bosons are considered. Finally, it is noticed that the infrared degrees of freedom of the half boson are protected against a particular coupling to noise in the model (this is an absence of Anderson localization for the low frequency modes). In section \ref{sec:spin} more general nonlinear models are considered. It is shown that the Toda chain and some of its generalizations can be constructed directly in this framework. If one considers periodic identifications of the bosonic degrees of freedom, one find systems with a reduced algebra that described spin chains with finitely many qubits per unit length. These turn out to be non-trivial critical spin chains that include the Ising model in a transverse magnetic field at criticality. The critical 3 state Potts model and some other systems with conserved ${\mathbb Z}_n$ charges. Higher dimensional generalizations are considered and the simplest model is shown to lead to a gapless fracton system.
 I then conclude.

\section{The chiral boson}\label{sec:chb}

The purpose of this section is to start with a simple example of a continuous field theory that describes a half boson: the chiral boson in $1+1$ dimensions. 
This model  arises naturally in many areas of physics like the Quantum Hall effect \cite{Wen:1989mw,Stone:1990iw},  and even as sectors in the gauge/gravity duality \cite{Berenstein:2004kk}.
The idea here is to motivate the precise structure of a Poisson bracket structure that will arise in lattice examples later on. 

Consider a field theory of a single chiral boson in one dimension. The equations of motion read $\partial_t \phi = c\partial_x \phi$ with $c$ the speed of light. 
The most general solution of  the partial differential equation is $\phi(x,t)=\phi(x+ct)$ and this represents a degree of freedom of a boson moving to the left at a fixed speed, which we call the speed of light. 
By contrast, a regular massless boson would have degrees of freedom that satisfy the second order wave equation
\begin{equation}
\partial_t^2 \phi = c^2 \partial_x^2 \phi .\label{eq:boson}
\end{equation}
The regular boson is a linear combination of a left moving boson and a right moving boson: we can split the solutions of the equation of motion into left and right movers and use
only half of the solutions. In a certain sense, we can say that a chiral boson is only half a boson degree of freedom rather than a full boson. 
The full boson degree of freedom has a simple lagrangian from which the equations of motion can be derived
\begin{equation}
L= \int dx \left[\frac{1}2 (\partial_t\phi)^2-\frac 12 (\partial_ x \phi)^2\right].\label{eq:KleinGordon}
\end{equation}
We can ask the question if there is a Lagrangian that produces the chiral boson equations of motion starting from $\phi$ alone, without using any additional fields? 
A slightly different formulation is the following: the Hamiltonian formulation of \eqref{eq:KleinGordon} has two local degrees of freedom: $\phi(x)$ and its canonical conjugate
$\pi(x)$, with local commutation relations $\{\phi(x), \pi(x')\} = \delta(x-x')$. Is there a Hamiltonian formulation that uses only a half of the degrees of freedom of the regular boson and
produces the correct dynamics?

It turns out that there is such a description. In that case we also need to consider the Poisson bracket of the fields $\{\phi(x), \phi(x')\}$, which should vanish for spatially separated $x, x'$, should be local and also antisymmetric. 
A Poisson bracket that satisfies the required antisymmetry properties is the following 
\begin{equation}
\{\phi(x), \phi(x')\}=\partial_x \delta(x-x') = - \partial_{x'}\delta(x-x'). \label{eq:PBphi}
\end{equation}
This is a classical form of a $U(1)$ Kac-Moody algebra, where  the boson $\phi \simeq J$ is the classical field associated to the current. The $\delta'$ commutation is the anomaly of the $U(1)$ current algebra.

On general functionals $F(\phi), G(\phi)$ we write their Poisson brackets as follows
\begin{equation}
\{F,G\} = \int dx d x' \frac{\delta F}{\delta \phi(x)} \frac{\delta G}{\delta \phi(x')} \{\phi(x), \phi(x')\} 
\end{equation}

The  Hamiltonian for the chiral boson is given by
\begin{equation}
H = \frac 12 \int c \phi(x)^2 dx, \label{eq:chb}
\end{equation}
we then find that the equation of motion is
\begin{eqnarray}
\dot \phi(x) &=& \{\phi(x), H\} = - c \int \phi(x')  \partial_{x'} \delta(x-x')  dx' \nonumber\\
&=& c\int \partial_{x'} \phi(x') \delta(x-x') = c\partial_x \phi(x),\label{eq:chbeom}
\end{eqnarray}
where to go from the top line to the bottom line we integrate by parts. 

Classical Poisson structures of these type arise from direct computation in both Quantum Hall \cite{Polychronakos:2004es} and gravity examples \cite{Maoz:2005nk} (these are very useful for the semiclassical description of excitations around a background field configuration \cite{Berenstein:2017rrx}).

There are now some interesting features that the degrees of freedom of \eqref{eq:PBphi},\eqref{eq:chb} satisfy that the corresponding full boson in \eqref{eq:KleinGordon} doesn't. The main feature I want to highlight is that the full boson degree of freedom admits a local relevant deformation, quadratic in the fields, that
gaps the system. This is the mass deformation. For the chiral half boson, the Hamiltonian is already the most relevant quadratic function of the fields $\phi$. It admits no such deformation. We also restrict ourselves to the algebra of the $\phi$ directly. 

There are many ways to argue that the chiral boson is protected. The simplest one is to argue that the chiral boson has a central charge $(c_L,c_R)=(1,0)$ that produces a gravitational  anomaly. Anomaly matching to the infrared prevents the bosons from developing a mass. 
This is different than a protection mechanism for \eqref{eq:KleinGordon} that would require $\phi$ to be a Goldstone boson for a spontaneously broken symmetry. The anomaly  is more robust.

Now, we find that the system that leads to \eqref{eq:chbeom} has a first order differential equation of motion. This is very  similar to fermion degrees of freedom that satisfy a Dirac equation, rather than bosons. 
The natural question that we will ask is if there is a discretized version of the Poisson bracket \eqref{eq:PBphi}, which only has one degree of freedom $\phi(n)$ at each site but no natural notion of $\pi(n)$ at the same site.
This is the model we will explore in the next section. After that model is found, one can ask what similarities with the chiral bosons survive the discretization and if there are new features that are intrinsic to this discretization that make it more interesting in its own right.

\section{The model}\label{sec:model}

When we have a boson at each lattice site, we usually have two canonical conjugate variables $x_i,p_i$ at each site (these are usually, $x$, $\dot x$ in a Lagrangian formalism), with Poisson bracket $\{x,p\}= 1$. A local Hamiltonian will then be a sum of terms where the variables $x_i,p_i$ appear in terms that include  only a finite number of nearby sites.
In the Hamiltonian formalism, $x_i,p_i$ can be treated as equivalent types of variables: one can always do a canonical transformation $x_i\to p_i, p_i\to -x_i$ that reveals that it is only the labels $x,p$ that distinguish them. Obviously, if the Hamiltonian has a special form, the $x,p$ might also be distinguished by how they appear in the Hamiltonian. This is a property of the dynamics under consideration, not the Poisson bracket itself.

These variables  are not intrinsically distinct as mathematical objects, so we can label them by $q_{i,1}, q_{i,2}$ and forget that one is called a position and the other one is called a momentum.  We think of them as a set of phase space variables instead, without any additional interpretation.

The full Poisson bracket is then  
\begin{equation}
\{ q_{i,\alpha}, q_{j, \beta}\} = \delta_{ij} \epsilon_{\alpha,\beta},\label{eq:fullboson}
\end{equation} 
where $\epsilon_{\alpha\beta}$ is the totally antisymmetric 
tensor in two dimensions.

We want to go to a staggered model where the notion of the variables $x,p$ become distributed over a lattice and only half of them survive at a lattice site, giving a half boson rather than a full boson degree of freedom.  
What does it mean to have a half boson per lattice site? 
We should think of this property as only having a single real $q_i$ variable per site, rather than two. Obviously, any Poisson bracket $\{q_i,q_i\}=0$ must vanish because of antisymmetry. 
A naive generalization of \eqref{eq:fullboson} with bosons that commute at different sites would vanish identically. Such a system would not evolve in time under Hamiltonian equations of motion.

However, we can consider a more general Poisson bracket that is not ultralocal: we  can ask that $\{q_i, q_j\}= \Omega_{ij}$  is an antisymmetric constant tensor, with non vanishing elements up to some set of nearby sites. In the continuum limit, such terms would become local. The simplest such Poisson bracket  that is also translation invariant is given by
\begin{equation}
\{q_i, q_j\} = \delta_{i,j-1}- \delta_{i,j+1}.\label{eq:halfboson}
\end{equation}
In this way, $q_{i+1}$ is like a momentum for $q_i$.
 There is a second model where we change all the signs. We will call the one above {\em left moving}, and the other one {\em right moving}.  We will see that these are equivalent in the end.
 This nomenclature will become more obvious after we write a Hamiltonian.
 This is similar to a discretization of the derivative of the Dirac delta function that we encountered in the previous section in \eqref{eq:PBphi}.
  
Consider the following quadratic Hamiltonian inspired by the chiral boson \eqref{eq:chb}
 \begin{equation}
 H = \sum_i \frac 12 q_i^2,\label{eq:model1}
 \end{equation}
 with the most local form possible and that is also translation invariant and a positive function: it is a sum of squares.
 
The equations of motion can be readily found
\begin{equation}
\dot q_n= \{ q_i, H\} = q_{n+1}-q_{n-1}. \label{eq:nearn}
\end{equation}
We see that the Poisson bracket itself is providing a dynamical coupling between neighbors. This is in contrast to other models where the coupling to the neighbors arises from the specific form of the Hamiltonian. 
In the equation \eqref{eq:nearn} we get an equation of motion that looks like a 
discrete version of the derivative. A naive continuum limit would then give us
\begin{equation}
\dot \phi = 2 \partial_x\phi, \label{eq:leftmover}
\end{equation}
which is a left chiral boson with speed of light $c=2$. It is this property that makes the naming convention for the {\em left moving} Poisson bracket. The factor of $2$ arises due to conventions
$q_{n+1}-q_{n-1} \simeq q(x+1)-q(x-1) \sim  2 \partial_x q$ for slowly varying bosons and plays the role of the speed of light.
The main idea here is that in the continuum long wavelength approximation the system becomes relativistic. This feature arose automatically from the Poisson bracket without any fine tuning. This suggests that the dynamics enjoys some form of protection of the infrared dynamics that distinguishes it from other
 field theories. 
 A lot of this paper is devoted to understanding this physics.

\subsection{The dispersion relation}

The system that was presented above is translation invariant and has linear equations of motion. It also has a conserved energy (the Hamiltonian)  that is bounded from below. The system is stable and should therefore admit a decomposition into harmonic oscillators where the solutions are plane waves. Let us write a possible such plane wave solution 
\begin{equation}
q_n = \exp( i k n - i \omega t) .
\end{equation}
Our goal is compute the dispersion relation $\omega(k)$. Putting this ans\"atz in \eqref{eq:nearn} we find that 
\begin{equation}
-i \omega =  \exp(i k)- \exp(-i k),
\end{equation}
so that 
\begin{equation}
\omega(k) = -2\sin(k) .\label{eq:nosqroot}
\end{equation}
Notice that there are two zeros: one at $k=0$ and the other one at $k=\pi$. The quasimomentum $k$ is periodic with period $2\pi$. The left mover is at $k=0$, and it is left moving because the group velocity is negative $\partial_k \omega|_{k=0}=-2$. Similarly, at $k=\pi$ we find a second zero of the dispersion relation. That one is a right mover, as $\partial_k\omega|_{k=\pi}=2$. 

There should be no surprise that the model actually has a right mover. The Nielsen-Ninomiya theorem \cite{Nielsen:1981hk} as relates to the dispersion relation that is polynomial (or smooth) predicts doublers.
Basically, if $\omega(k)$ is periodic, continuous and if there is a crossing of zero at some value, there must be another crossing at some other value.
This is as true for bosons as it is for fermions. Alternatively, one can say that since in the UV there is a lattice cutoff, the UV central charge $c_L-c_R$ must vanish (there are no degrees of freedom beyond the cutoff scale). Anomaly matching then predict that if there is a sector with $c_L=1$, there must be an accompanying sector with $c_R=1$ to cancel the gravitational anomaly (see also the discussion in \cite{Hellerman:2021fla}, which shows that any lattice flow must have a nontrivial conformal boundary state).

What is interesting about this model is that in equation \eqref{eq:nosqroot} there is no square root appearing in the dispersion relation: $\omega(k)$ is a single valued function of $k$, rather than a double valued function of $k$, like one would obtain for a regular boson. 
This suggests that the gap can not be lifted by the following topological argument: any small deformation of $\omega(k)$ needs to maintain the crossings of zero as in some regions $\omega(k)>0$ and in others $\omega(k)<0$. The left and right moving bosons are located at these crossings: topology in the dispersion relation prevents the gap from forming and the left and right movers from mixing with each other in the infrared: they are located at different values of $k$ so they cannot scatter into each other. 

Let us now make the right mover more explicit. Consider a new set of variables $\tilde q_i =(-1)^i q_i$. If we work with $\tilde q$ instead of the $q$, the Poisson bracket changes sign (the nearest neighbors of the even sites get a sign change, whereas the even variables do not). As we see the {\em right mover} is also part of the discrete model of the half boson. The naive continuum expressed in \eqref{eq:leftmover} fails only because it implicitly assumes that $q$ does not change very much from neighbor to neighbor. A better way to take the continuum is to assume that we have both a slowly varying $q$ superposed with a slowly varying $\tilde q$: we keep all the oscillator degrees of freedom that give rise to a small $\omega(k)$, no matter what the value of $k$ is. The slowly varying $\tilde q$ is Neel-ordered relative to the slowly varying $q$ functions in the $q$ variables, so they are independent degrees of freedom in the IR (small $\omega$ limit, as opposed to small $k$). 

In this way we have both the left mover and the right mover bosons appear as independent infrared continuum degrees of freedom.

\section{Topology}\label{sec:top}

So far, I have described a simple model of dynamics on a lattice, where at each site there is a single bosonic degree of freedom, rather than a pair of conjugate degrees of freedom.  The Poisson bracket is described by a nearest neighbor non-trivial commutation relation. I will now show that this model predicts various interesting features that can be associated with topological features. First, the half boson stays massless if we deform the quadratic Hamiltonian and keep translation invariance. Essentially, the left and right movers are not allowed to hybridize.  I will also show that the model I described above has a non-trivial parity symmetry and I will explore its consequences.

\subsection{Protected massless bosons}

As I have shown, the simplest model of half bosons produces a gapless quantum field theory in the infrared. I want to show that this result extends to the most general stable quadratic Hamiltonians that are local up to $s$ nearest neighbors. Essentially, this gives a proof that the left and right movers are not allowed to hybridize if we preserve translation invariance. The argument is rather simple. A term in the Hamiltonian including $q_i$ would have terms with up to $q_{i-s}, \dots, q_{i+s}$, of the general form
\begin{equation}
q_i( \sum_{t=-s}^s a_{t} q_{i+t}) .
\end{equation}
where the $a_t$ are real, as we are assuming that the $q$ are real variables. It is also the case that $a_t= a_{-t}$, as both of these arise from the same term $q_{k} q_{k+t}$ for $k=i$, or $k=i-t$.  We also need some positivity condition to be determined later.

The equation of motion follows:
\begin{equation}
\dot q_n = ( \sum_{t=-s}^s a_{t} q_{n+t+1}) -( \sum_{t=-s}^s a_{t} q_{n+t-1}) .
\end{equation} 
Now we solve for the dispersion relation by using the plane wave ans\"atz, $q_n = \exp( i k n - i \omega t) $ to get 
\begin{eqnarray}
-i \omega &=& ( \sum_{t=-s}^s a_{t} (\exp (ik(t+1))-\exp(ik(t-1) )\\
&=& 2 i \sin(k)  \left(a_0 +\sum_{t=1}^s 2 a_{t} \cos( k t)\right)  .
\end{eqnarray}
This shows that there are always zeros of the dispersion relation at $k=0, \pi$, no matter what we do. This is slightly stronger than the continuity argument presented in the previous section: indeed, these are the only zeros and they have fixed location. The two zeros at $k=0$ and $k=\pi$ can 
not hybridize as they are at a different value of the quasimomentum. Basically, they have different eigenvalues with respect to translation symmetry. This means that the massless modes are protected by symmetry considerations.

Finally, the positivity condition we need is that the function 
\begin{equation}
f(k)= \left(a_0 +\sum_{t=1}^s 2 a_{t} \cos( k t)\right)> 0, \label{eq:dispgen}
\end{equation}  for all $k$. In particular this implies that $a_0>0$, so that the
Hamiltonian can be considered as a  {\em continuous} deformation of the one presented in the previous section. 

The most natural way to write the Hamiltonian is to go directly to a momentum basis. Let us expand the $q_x$ as follows
\begin{equation}
q_x = \int_0^{2\pi} \frac{dk}{2\pi} \exp( i kx) c_k.
\end{equation} 
Inverting the Fourier transform, we get that 
\begin{equation}
c_k= \sum_x q_x \exp(-i kx).
\end{equation}
When we compute the Poisson brackets of these, we get that 
\begin{eqnarray}
\{c_k, c_{\tilde k}\} &=& \sum_{x,x'} \exp( -i k x) \exp( -i\tilde k x') (\delta_{x,x'+1}-\delta_{x,x'-1})\nonumber\\
&=&  \sum_{x} \exp(-ik x-i \tilde k x)  (\exp(-i\tilde k)-\exp(i \tilde k) ).\nonumber\\
&\equiv& \delta(k+\tilde k) (2 i) \sin(k) \label{eq:normalization}
\end{eqnarray}
If we quantize, we find that the $c$ mode at $k$ and the one at $(-k)$  become raising and lowering operators. Basically, quantizing replaces Poisson brackets by commutators, with an extra factor of $i$.
In the quantization, the extra factor of $i$ appearing on the right hand side  disappears.

Which $a_k, a_{-k}$ acts as a  lowering operator and which one is acting as a raising operator is completely determined by the sign of $\sin(k)$ in the commutation relations. 
  The Hamiltonian is then proportional to
\begin{equation}
H \propto \int_0^{2\pi} dk f(k) c_k c_{-k}. \label{eq:ham_gen}
\end{equation}
with $f(k)$ real and polynomial in $\exp(ik)$ as described in \eqref{eq:dispgen}.
It is  positive if $f(k)>0$, to ensure that the ground state with all oscillators at zero occupancy has the minimal energy. It is very important to notice that the $c$ operators are not canonically normalized. They are instead subject to \eqref{eq:normalization}.

If we compare this Hamiltonian to a more regular boson, the dispersion relation of the general boson is instead of the more general form $\omega =\pm \sqrt{f(k)} $. 
The double valuednes allows one to have $f(k)>0$ everywhere: there is both a raising and a lowering operator for each value of $k$. 
In the half boson case described in this paper, it is either only a single raising operator or only a single lowering operator at each value of $k$. The conjugate variable to the one labeled by $k$ is still at $-k$, however. The points of symmetry of the operation exchanging $k\to -k$, namely $k=0,\pi$ (remember that $k$ is periodic with modulus $2\pi$) are self conjugate: 
they are neither raising nor lowering operators. Instead, they are central elements in the algebra of raising/lowering operators: they commute with everything. This means that they can't evolve with $H$ and therefore must get stuck at zero frequency. 

\subsubsection{Parity symmetry}

The quadratic simple model has a parity symmetry. This is not obvious at first. The idea is that $q_n \to q_{-n}$ turns the left mover into a right-mover. We fix this by using the $q_n\to (-1)^n q_{n}$ transformation that turns a right mover Hamiltonian back into a 
left moving presentation. The consequence of this parity symmetry can be better expressed in terms of the $c_k$ modes. It turns
$c_k \to c_{\pi-k}$, where $\pi-k$ is evaluated modulo $2\pi$. When we consider equation \eqref{eq:ham_gen}, we find that the parity symmetry survives if $f(k)=f(\pi-k)$. This condition is equivalent to 
\begin{equation}
\left(a_0 +\sum_{t=1}^s 2 a_{t} \cos( k t)\right)= \left(a_0 +\sum_{t=1}^s 2 a_{t} \cos( (\pi-k) t)\right).
\end{equation}
The two Fourier sums are the same if $a_t=0$ for $t$ odd. That is, the parity symmetry survives if we only have interactions between even-even sites and odd-odd sites, but no mixing between them. 
When the parity symmetry is present, the infrared dispersion relation of the left movers and the right movers gives the same IR speed of propagation. In that case, the infrared physics is universal and independent of most of the details.
It coincides with a free boson, unless one engineers the system so that the speed of light in the infrared vanishes. Essentially, there are no relevant translation invariant quadratic deformations of the infrared physics associated to the $q$ variables that also preserve the parity symmetry.
The most they can do is change the speed of light, which can be undone by a rescaling of the units of time. 

\subsection{Exact zero modes on finite lattices}

So far I have described the infinite volume limit of the half boson lattice field theory. I will now consider what occurs at finite volume with a periodic lattice of length $L$. Let us consider the dispersion relation \eqref{eq:nosqroot}. 
The quantization of $k = 2\pi s /L$ produces one zero mode if $L$ is odd (at $s=0$) and two zeros if $L$ is even (at $s=0, L/2$). 

At $L$ large but finite, the spectrum of the left mover has energies $\omega_s= 2 *(2\pi s)/L$ for integers $s$,  $s\geq 0$. Basically, we get a left boson with periodic boundary conditions. 
For $L$ even, the parity operator described previously is  a symmetry of the system and we get a similar spectrum of right movers, also with periodic boundary conditions. 
When $L$ is odd, the region near $s\simeq L/2$ misses the zero by a half integer unit. The spectrum is therefore $\omega_s= 2 *(2\pi (s+1/2))/L$ which gives anti-periodic boundary conditions for the right moving 
boson.

Here, there is a slight distinction between the right moving and the left moving Hamiltonians: the right moving Hamiltonian will produce a zero mode for the right movers on a lattice with an odd number of sites and the left moving Hamiltonian
will instead produce a zero mode for the left moving modes. The right moving Hamiltonian is equivalent to the left moving Hamiltonian with {\em anti-periodic boundary conditions} in an odd lattice. This is what takes care of the sign difference when going around the circle. 

 Basically, there is a topological property of the UV lattice that survives all the way to the infrared: if the lattice has an even or an odd number of sites. 
 This property determines if the total number of zero modes is even or odd, regardless of the boundary conditions being periodic or anti-periodic.

 This property can be traced back to the symplectic form $\Omega_{ij}$, which is an antisymmetric real matrix. The spectrum of eigenvalues of $\Omega$ is such that the eigenvalues are paired, with opposite signs.
 If there is an odd number of eigenvalues, then one of them must necessarily vanish. If there is an even number of eigenvalues, the number of zero eigenvalues of $\Omega$ must be even. 
Also, there are sectors with two zero modes (periodic boundary conditions on a lattice of size $2L$), or with no zero modes (antiperiodic boundary conditions on a lattice of size $2L$). 
 This is somewhat reminiscent of the problems of spin structures in string theory: if one has Ramond or Nevwu-Schwarz boundary conditions for left movers and right movers separately from each other (see \cite{Polchinski:1998rr}, chapter 10 for a textbook discussion). 

We can also consider the same problem on an interval (open boundary conditions). It is easy to check that we can always turn a left moving Hamiltonian into a right mover in this case, as the positions on the lattice are ordered and we do not have to worry about how the phase winds around the circle. 

To the extent that there is a zero mode, it is equally shared between the left and right movers. 
When the lattice has an odd number of sites, there is a zero mode, by the same argument on $\Omega_{ij}$ given above. If the sites go from $0, \dots, 2k$, the zero mode is 
$c_0= q_0+\dots + q_{2k}$ which can be checked to commute with all other $q$. This lis similar to having a differential equation with Neumann-Neumann boundary conditions. 
One can check that there is no zero mode if the  number of sites is even. This is similar to having problems with Dirichlet-Neumann boundary conditions \footnote{Both types of mixed boundary problems arise naturally in string theory when discussing intersecting D-branes \cite{Polchinski:1995mt}}.  
This phenomenology is reminiscent of the Majoprana fermions studied by Kitaev \cite{Kitaev:2000nmw}, where the number of zero modes depends on if there is an even or an odd number of sites.

\subsection{Projections from  regular bosons to  half bosons}

In what sense is a regular lattice boson a combination of two half bosons? More precisely, it is interesting to ask the question in reverse: can one construct a half boson from a projection of a regular boson?
I will answer that question in the affirmative. This will also explain some issues with zero modes as to what is missing when we combine half bosons together.

The idea is to start with regular bosons at lattice sites $x_i, p_i$, with $\{x_i, p_j\}=\delta_{ij}$. I want to generate a non-trivial nearest neighbor Poisson bracket. The $q$ variables need to be linear combinations of the $x,p$. To do that, consider the following expression:
\begin{equation}
q_i= p_i+x_{i+1}.
\end{equation}
One easily finds that
\begin{equation}
\{q_i, q_{i+1}\} =\{x_{i+1}, p_{i+1}\} =1 ,
\end{equation}
where we see that there is a perfect match with equation \eqref{eq:halfboson}, so this is a half boson. 
Consider also $w_i = x_{i}+p_{i+1}$. It's easy to show that these two sets of variables commute $\{w_i, q_s\}=0$.
Moreover
\begin{equation}
\{w_i, w_{i+1}\} = - 1 
\end{equation}
This shows that in the theory of a full boson, we find two half bosons that commute with each other. One of them is naturally a left mover and the other
one is naturally a right mover.  
It is also easy to check that $c_0=\sum_i q_i = \sum_i w_i$, so the central elements of the zero modes match between these. This also applies to the zero mode at half momentum of the variables $\tilde q_i$ and $\tilde w_i$ that one could obtain, with a difference in sign.
Notice that on the other hand, the $x,p$ variables have a non-degenerate Poisson bracket. What this means is that the projection of the boson to the two half bosons leaves something out. 
What are ctually missing  the conjugate variables to $c_0, c_\pi$. 

What this means is that if we combine two half bosons together, we will not get a full boson. At best, the zero modes of the half bosons are shared, but some modes will be missing that  are conjugate to the 
zero modes. This explains why even though we are able to hybridize the massless full bosons to remove the massless modes, this will not be possible if we have two half bosons stitched together. The half boson zero modes 
will persist: their conjugate variables are not part of the half bosons algebras. We can say that the topological protection of the half bosons is due to this absence, even in the presence of other half boson modes. 
The continuity argument of the dispersion relation would not forbid two half bosons combining to a full boson and gapping the theory. 

There is a second projection we can consider: from a boson on a lattice to a half boson on a lattice that is twice as large. This would correspond to the $2N$, $x,p$ variables, being mapped to $2N$ $q$ variables.
The idea is then to take
\begin{eqnarray}
q_{2i}&=& p_i \nonumber\\
q_{2i+1}&=& x_{i+1}- x_i .\label{eq:singleb}
\end{eqnarray}
One can easily check the commutation relations. The Hamiltonian is invariant under shifts of $x$, with all $x$ shifted simultaneously. This would naturally be associated to a shift symmetry being present and part of the low energy Lagrangian. Since the shift symmetry is non-linearly realized, one expects a Goldstone boson.

It is easy to check that the Hamiltonian is parity invariant. Translation invariance on this lattice is more subtle.
The Poisson brackets relations are translation invariant. However, in this projection 
there is only one zero mode that is equally shared between the left movers and right movers. One can easily check that the sum $\sum q_i = \sum q_{2i}= \sum p_i$ 
is the only zero mode.  There is one zero mode linear combination that is identically equal to zero if we insist that the $x$ variables (rather than the $q$ variables) are periodic in $L$. 
In the theory of the $x,p$ variables, the  zero mode that survives is the center of mass momentum. The variables $q$ are relative positions, so the center of mass coordinate is missing.

Apart from the zero mode, the theory is translation invariant. If we take two steps, the theory is translation invariant $mod(2)$. This would be the ordinary translation invariance of the lattice of the $x,p$ variables. One can say that the topological protection that the system enjoys has arisen from a  dynamical square root of the translation operator that is present in the system. This is in lieu of saying that one has a shift symmetry of all the $x$ variables simultaneously (the Goldstone boson argument). 

We can also insist on preserving the zero mode of the $q$. 
From the point of view of the $x$ variables, we find that the 
sectors with the extra zero mode turned on differ by a shift when one considers going around the circle in the $x$ variables. One needs to solve the difference equation $x_{i+1}-x_i=const$ consistent with a periodic condition on the  $q$. 

This is akin to winding modes for a boson theory (see \cite{Polchinski:1998rq}). When both winding and momentum appear, one can have symmetries that exchange them. 
This is called T-duality in string theory. What we find is an instance where the $q$ variables at the classical level can lead to both momentum and winding being continuous variables. There is a certain sense in which the classical theory that appears in the way I have described cannot distinguish the two notions. 
When quantizing, only some combinations of momentum and winding that are related to each other in particular ways lead to modular invariant partition functions (this is the problem studied more generally in \cite{Narain:1985jj}).
At this stage, there is no requirement of having a modular invariant answer for the partition function in the infrared. After all, the formulation of this dynamics was done directly in a real time first order formulation of the theory. One cannot immediately assume that there is  a Euclidean partition function that should satisfy modularity.
This is similar to the chiral boson: it is a consistent relativistic theory in one dimension, but it does not have a modular invariant partition function.

\subsection{Coupling to noise and absence of Anderson localization in the IR}

Let us now consider the problem of coupling the half bosons to noise. The idea is to consider a Hamiltonian of the form
\begin{equation}
H= \sum\frac {\eta(i)}{2} q_i^2,  \label{eq:noise}
\end{equation}
where $\eta$  is a positive random variable. For simplicity,  a model where $\eta$ will be chosen uniformly randomly between $0.7$ and $1$ is considered. This is
a version of noise. A natural question is if the modes of the system exhibit Anderson localization \cite{Anderson:1958vr} or not. In a sense, this is another way to check if the phase of matter associated to
the staggered bosons is robust to noise or not. 

\begin{figure}[ht]
\includegraphics[width=7cm]{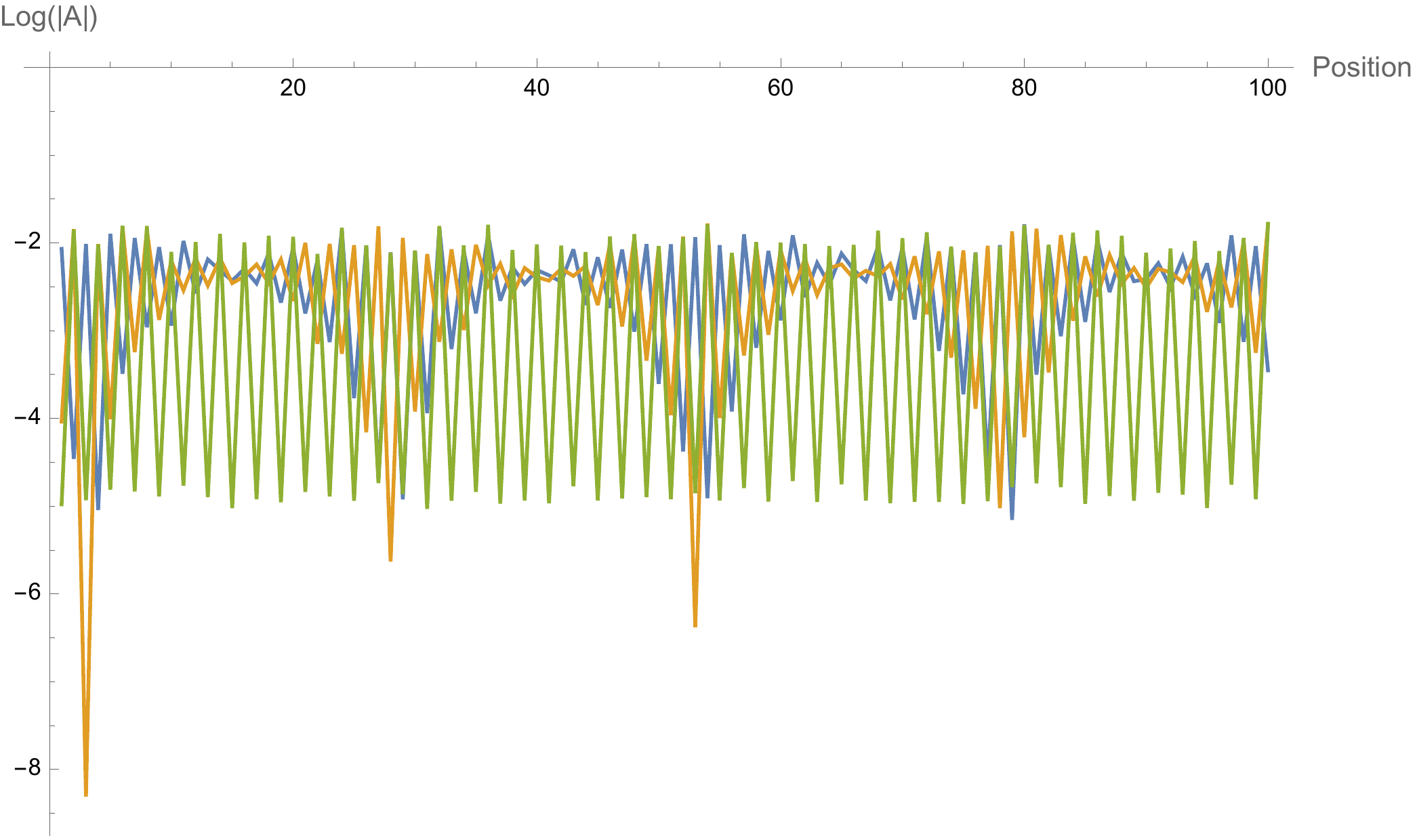}\includegraphics[width=7cm]{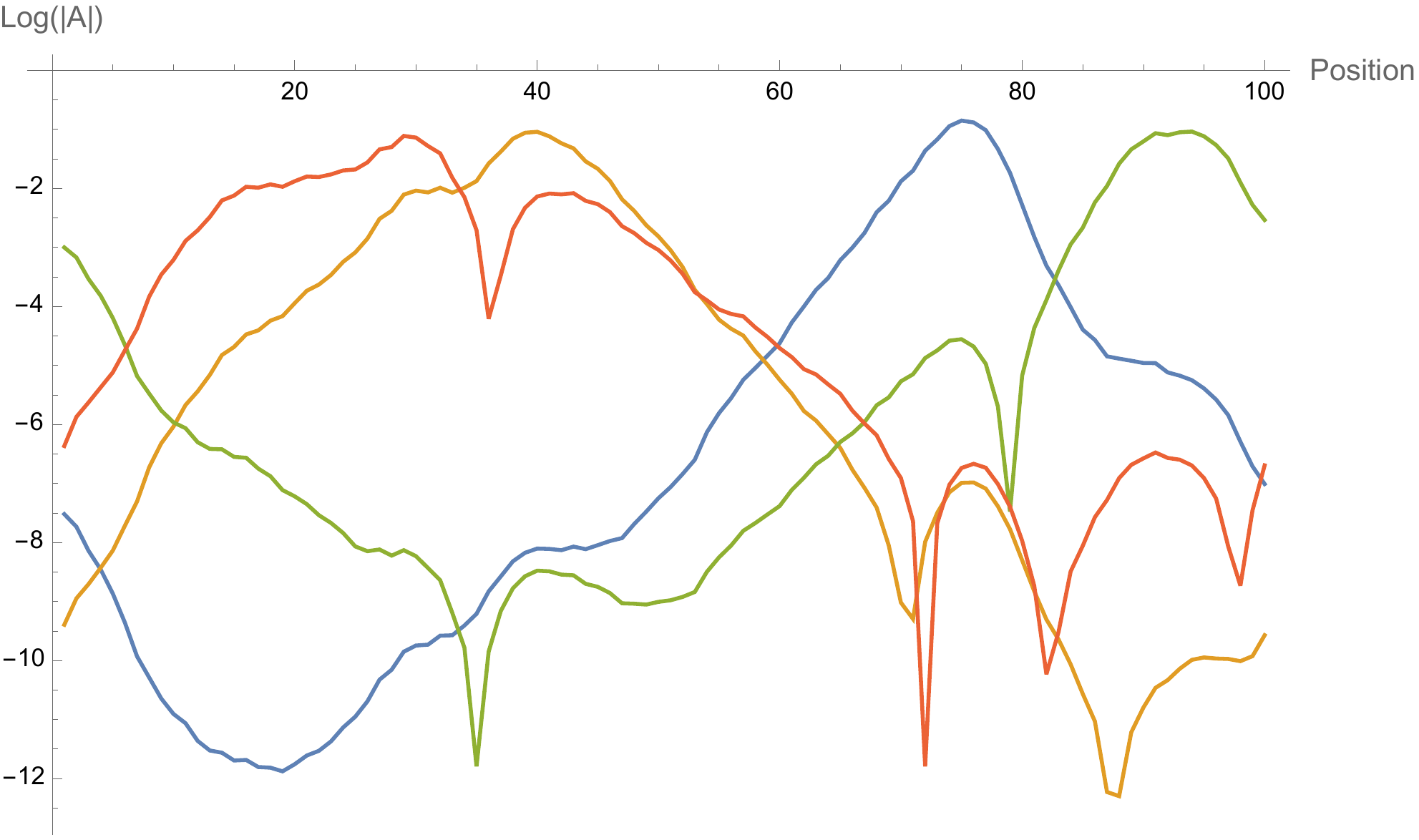}
\caption{Logarithm of position amplitudes of modes from a model with 100 sites are shown.
On the left, we consider the normal modes near zero frequency. On the right we do the same for the highest frequency modes}
\label{fig:loc}
\end{figure}
This is checked this numerically in figure \ref{fig:loc}. On the left, we see that the size of the  amplitudes of the low lying modes stay fairly constant on the lattice at long distances. On the right figure we see that for high energy modes, the modes decay very quickly from a position with high amplitude towards regions that are far away in the lattice. The high energy modes (that in some sense are irrelevant for the infrared description of the theory) display Anderson localization. By contrast, the low energy modes do not.
The goal right now is to explain this lack of localization on the long wavelength excitations. This is some indication of  robustness of the staggered boson theory, similar to topological order in quantum Hall systems that forbid the modes from getting lifted by defects. 
 A way to see what is going on is to actually look at the numerical value of the low lying eigenvalues determining the frequencies of the low lying modes.

\begin{figure}[ht]
\includegraphics[width=8cm]{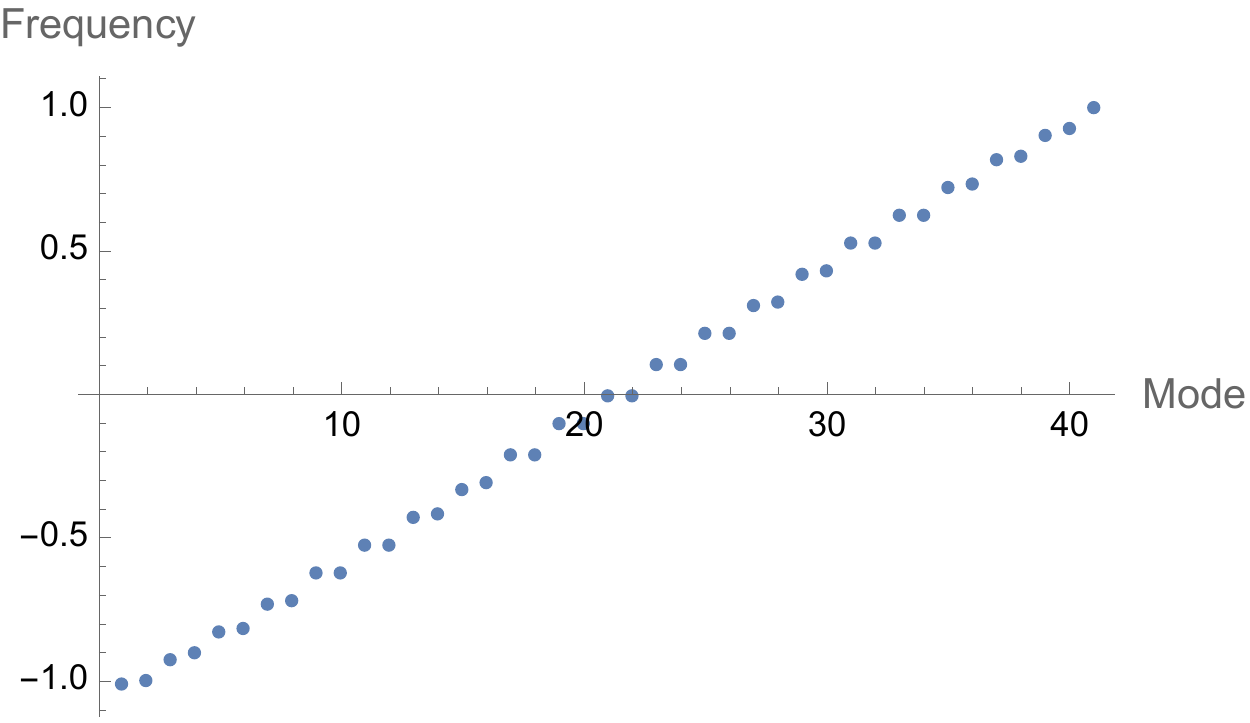}
\caption{The numerical value of the frequencies of eigenvalues near zero energy is shown}
\label{fig:spec}
\end{figure}

 This is shown in figure \ref{fig:spec}. 
What is seen are two exact zero modes (the same zero modes we have been discussing above) and a spectrum that is nearly doubly degenerate at each value, as well as evenly spaced.
Such a spectrum would arise from a Conformal Field Theory on a circle: a left moving and a right moving oscillator for each Fourier mode on a circle.

I want to explain this result more intuitively. The idea is that the model described by the Hamiltonian \eqref{eq:model1} is gapless in the infrared and leads to a $c=1$ boson. 
In such  system, we would have both a left current density $J \simeq \partial_+\chi$ and right handed one $\bar J \simeq \partial_- \chi$ and the Hamiltonian would be of the Sugawara form
\begin{equation}
H \simeq \int dx \left[\frac 12 :J(x) J(x):+\frac 12 :\bar J(x) \bar J(x):\right].\label{eq:sug}
\end{equation}
The natural notion of the discretized current is $J_i\simeq q_i+q_{i+1}$, so that  it  couples mostly to the left movers. By contrast $\bar J_i\sim \tilde q_i+\tilde q_{i+1}$ couples mostly to the right movers. The variables $q$ alone are not slowly varying in the procedure we use to get to the infrared, but $J_i$ and $\bar J_i$ are. 
Because of the alternating signs between $\tilde q_i $ and $q_i$, one can check that the Hamiltonian \eqref{eq:model1} can also be written as sums of squares 
$H\propto (q_i+q_{i+1})^2+(\tilde q_i+\tilde q_{i+1})^2$. The cross terms cancel. Let us now turn to the noise model \eqref{eq:noise}. A natural way to think about it in the $J, \bar J$ variables would be to take the form
\begin{equation}
H \equiv \int   dx  \eta(x)\left[\frac 12 :J(x) J(x):+\frac 12 :\bar J(x) \bar J(x):\right],
\end{equation}
where we have added a field $\eta(x)$ that varies locally and has some noise. This is like coupling to a random metric on the circle, but keeping time translation invariance. 
With a change of coordinates in $x$, we should be able to bring the Hamiltonian back to the form \eqref{eq:sug} in the continuum limit. This means that the spectrum should be the one of a $c=1$ conformal field theory on a circle whose size depends on $\eta$. Also, since the system preserves the left-right symmetry of expressions in terms of $J, \bar J$ (when averaging over the $\eta$), the left and right moving frequencies should be approximately degenerate on any instance of the $\eta$. What we see is that the conformal symmetry persists in the infrared and the speed of light of left and right movers is the same.
By contrast, the UV modes are near the cutoff: they can resolve the lattice and do not average out the fluctuating lattice properties. Fluctuations in $\eta$ will be visible to these modes and  they can exhibit localization: they do so in the example above.

\section{Algebraic truncations and spin chains}\label{sec:spin}

It is interesting to study more general models than quadratic models in the $1-D$ lattice. To warm up, let us consider the following Hamiltonian, which only has translation invariance modulo $2$:
\begin{equation}
H = \sum_i \frac 12 q_{2i}^2 + \exp(q_{2i+1}).\label{eq:Toda}
\end{equation}
If we go back to the map from a single boson to the $q$ variables   \eqref{eq:singleb}, we see that we get the Hamiltonian for the Toda chain \cite{Toda}. 
We might worry that modifications of this Hamiltonian might lead to different dynamics, with for example
\begin{equation}
H = \sum_i \frac B 2 q_{2i}^2 + C \exp( \alpha q_{2i+1}).
\end{equation}
It is easy to check that the Poisson bracket is invariant under the following staggered rescaling  $q_{2i} \to \beta q_{2i}, q_{2i+1}\to \beta^{-1} q_{2i+1} $, so we can remove $\alpha$. Similarly, the zero mode of the odd $q$, $\sum_i q_{2i+1}$ lets us modify $C$ by a global translation of the odd $q$. Finally, if we change the units of time we can change $B$, so we can go back
 to \eqref{eq:Toda}. We can now write a translation invariant modification of the Toda chain as follows  
 \begin{equation}
\tilde  H = \sum_i \exp(\alpha q_{i}).\label{eq:Todaexp}
 \end{equation}
 which is similar to the relativistic Toda chain (one can also easily formulate the relativistic Toda chain Hamiltonian in the conventions of \cite{Hatsuda:2015qzx}).
 
Unlike the case \eqref{eq:Toda}, in this case the coupling  $\alpha $ is physical as we can not just change it by a staggered rescaling that keeps the Poisson  bracket structure intact while also preserving translation invariance.

In the classical limit this does not matter: we can effectively rescale the Poisson bracket by changing the units of time.  In the quantum case this cannot be done.
 We are therefore interested in the case where the variables $q$ are quantum variables rather than classical variables. The commutation relations in natural units $\hbar=1$ are then
 \begin{equation}
 [q_m, q_j] = i \delta_{m,j-1}- i \delta_{m,j+1}
 \end{equation}
 Now, we see that the new exponential Hamiltonian is generated by  new variables of the type $K_j = \exp(\alpha q_j)$. We want to understand the commutation relations of the $K_j$, rather than the $q_j$. We find, using the Baker-Campbell-Hausdorff formula that 
\begin{equation}
\exp( \alpha q_j) \exp(\alpha q_{j+1}) = \exp( \alpha (q_j+q_{j+1}) + i \alpha^2/2) =\exp(\alpha q_{j+1})\exp( \alpha q_j) \exp( i \alpha^2),
\end{equation}
so that 
\begin{equation}
K_j K_{j+1} = K_{j+1} K_j \exp( i \alpha^2).
\end{equation}
The commutation results in a phase. For certain values of $\alpha$ all of them commute with each other, or some powers of $K^m_j=\exp(m \alpha q_j)$ commute with the Hamiltonian. They would then be a conserved charge that appears quantum mechanically and is not immediately part of the classical conserved charges. 
This indicates that in the quantum case there can be a large center. 
This is suggestive that  the full Hamiltonian \eqref{eq:Todaexp} could be integrable, even in the classical limit. That problem is beyond the scope of the present work and will be investigated elsewhere.

The most interesting case to study is when $\alpha $ is imaginary, rather than real. The operator $\exp(i \alpha q)$ is actually a type of translation operator and is unitary, since $q$ is as a self-adjoint operator. 
Notice that these are invariant under translations of $q\to q+ 2\pi/\alpha$.  These are symmetries of the affine plane made of the $q$ variables that can be gauged. Basically, they preserve the Poisson brackets and the reality of the $q$ variables. 
We can therefore reduce the algebra of operators to those that are gauge invariant \footnote{We are not really allowed to gauge these  in the other case when the $\alpha $ quantity in the exponential is real rather than purely imaginary.}. When we gauge the translation symmetries, we find that the phase space becomes a $2L$ dimensional torus if we have $2L$ lattice sites. For the time being we will ignore how zero modes behave. 
 At the classical level, this doesn't affect much. 
Quantum mechanically the answer is different.  This becomes a phase space with finite volume and therefore it should be associated to a Hilbert space of finite dimension.
In that sense, one should be able to encode it into a finite number of qubits.

 If $\alpha$ is quantized appropriately, the quantity $\exp(i\alpha^2)$ is an $n-th$ root of unity.
This is a case where we have that the Hamiltonian is invariant  by finite translations in $q\to q+2\pi /\alpha$, and that the gauge invariant variables $K_j$ are such that $K_j^n$ is central. Since the center is large (commutes with the Hamiltonian), the corresponding variables can be chosen to be fixed 
(they can be diagonalized). That is, $\exp(i n \alpha q)$ 
is a c-number and the different eigenvalues of $K$ must be related to each other by roots of unity. 
The simplest gauge fixing is requiring that they are all frozen to the identity $K^n_j=1$.

This results in a  reduction of effective degrees of freedom.  The Hamiltonian \eqref{eq:Todaexp} would not be Hermitian if $\alpha$ is purely imaginary. 
This lack of Hermiticity can be remedied by adding the complex conjugate
\begin{equation}
H = -\sum_j  K_j+K_j^{-1} \simeq = -\sum _j 2 \cos( \alpha q_j) 
\end{equation}
where we insert a minus sign for convenience: after all, the $K$ are bounded operators, so the spectrum of $H$ is bounded both from above and below. The sign guarantees that the minimal energy of the classical state is at $q_j=0$.
When $\alpha \to 0$ (the periodicity is large), the Hamiltonian can be expanded  in Taylor series and we find that 
\begin{equation}
H \simeq \sum_j - 2 + \alpha^2 q_j^2- \frac 2{4!} (\alpha^4 q_j^4) +\dots
\end{equation}
So the quadratic piece is the same as the one of the simple model \eqref{eq:model1} and the quartic term is suppressed by $\alpha^2$ which is small at the cutoff scale. 
In that case perturbation theory should give a good approximation to the physics and the zeros of the dispersion relation we have discussed will stay in place: the model should be gapless at very small  $\alpha$ and have central charge $c=1$.
We can actually do better by noticing that the clock-shift matrices have similar root of unity commutation relations. In analogy with the map to the Toda chain, we should think of the even sites as having $\exp(i \alpha p)$ and the odd sites as $\exp( i \alpha \Delta x)$, as the map  \eqref{eq:singleb} would suggest.
The individual $\exp(i \alpha x)$, $\exp( i \alpha p)$ are magnetic translations and are realized by $P,Q$ clock shift matrices where $PQ= \omega Q P$ and $\omega^n=1$ and we add the constraint $P^n=Q^n=1$. These $P,Q$ are then given by  $n\times n$ matrices. That is, the Hilbert space of a single $P,Q$ pair is of dimension $n$. On a lattice of length $L$, these would given rise to a Hilbert space of size $n^L$. 
The map to a  spin chain  gives
\begin{eqnarray}
K_{2j+1} &\simeq& P_j \otimes P_{j+1}^{-1}\nonumber \\
K_{2j} &\simeq& Q_j
\end{eqnarray}
We can change left movers into right movers by changing $Q\to Q^{-1}$. 
In this setup, we have the Hamiltonian
\begin{equation}
H =- \sum_j \left[ Q_j+Q_j^{-1}+P_j \otimes P_{j+1}^{-1}+P^{-1}_j \otimes P_{j+1}\right]\label{eq:PQ}
\end{equation}
The terms $P\otimes P^{-1} +c.c$ can be interpreted as hopping terms, while the $Q+Q^{-1}$ are interpreted as a transverse field. The non-hermitian version of this model with the Hamiltonian
\begin{equation}
H =- \sum_j \left[ Q_j+P_j \otimes P_{j+1}^{-1}\right]\label{eq:Clock}
\end{equation} 
is actually a special case of the Baxter's clock model (see \cite{Fendley:2013snq}) for details.

The original model in terms of the $K$ variables lives on a lattice of size $2L$. The $P,Q$ model has a lattice of size $L$. There is a square root of the translation operator in the $Q,P$ lattice that effectively enhances the translation  symmetry, just like in the $x,p$ model. Just as in that case, we only have one zero mode rather than two. 
This is represented by the gauge invariant operator
\begin{equation}
\Omega= \prod_i Q_i\label{eq:charge}
\end{equation}
that commutes with Hamiltonian. 
It is such that $\Omega^n=1$ and it is a conserved charge modulo $n$.

If we consider the special case  of strong coupling $\omega^2=1$, the $P,Q$ matrices become Pauli matrices $P\simeq \sigma_z$, $Q\simeq \sigma_x$. In that case the Hamiltonian \eqref{eq:PQ} is exactly the critical Ising model in a transverse magnetic field. 
Surprisingly, the model is still gapless but the central charge is $c=1/2$.  It can be solved exactly by fermionization \cite{Fleury}.
The next special case is $\omega^3=1$, which gives the critical three state Potts model with $c=4/5$ ( see \cite{Wu:1982ra}).
At $\omega^4=1$, one seems to get to finally get to the central charge $c=1$  and that seems to be the case for all $\omega^n=1$, with $n\geq 4$. When $n\geq 5$ there seems to be states with the  conformal weights of $J, \bar J$ in the infrared limit, when computing on a lattice  with periodic boundary conditions.
One associates these to a BKT transition \cite{Berezinsky:1970fr,KT}.

 This is currently being investigated numerically \cite{BerTh}  by the same methods as in \cite{Milsted:2017csn}.  
These Hamiltonians are very similar to those that appear in \cite{Bazavov:2015kka} to study quantum implementations of the Abelian Higgs model in one dimension. They are also expected to lead to BKT transitions at some values of the couplings \cite{Zhang:2021dnz}.

\section{Higher dimensional models and Fractons}

\begin{figure}[ht]
\includegraphics[width=8 cm]{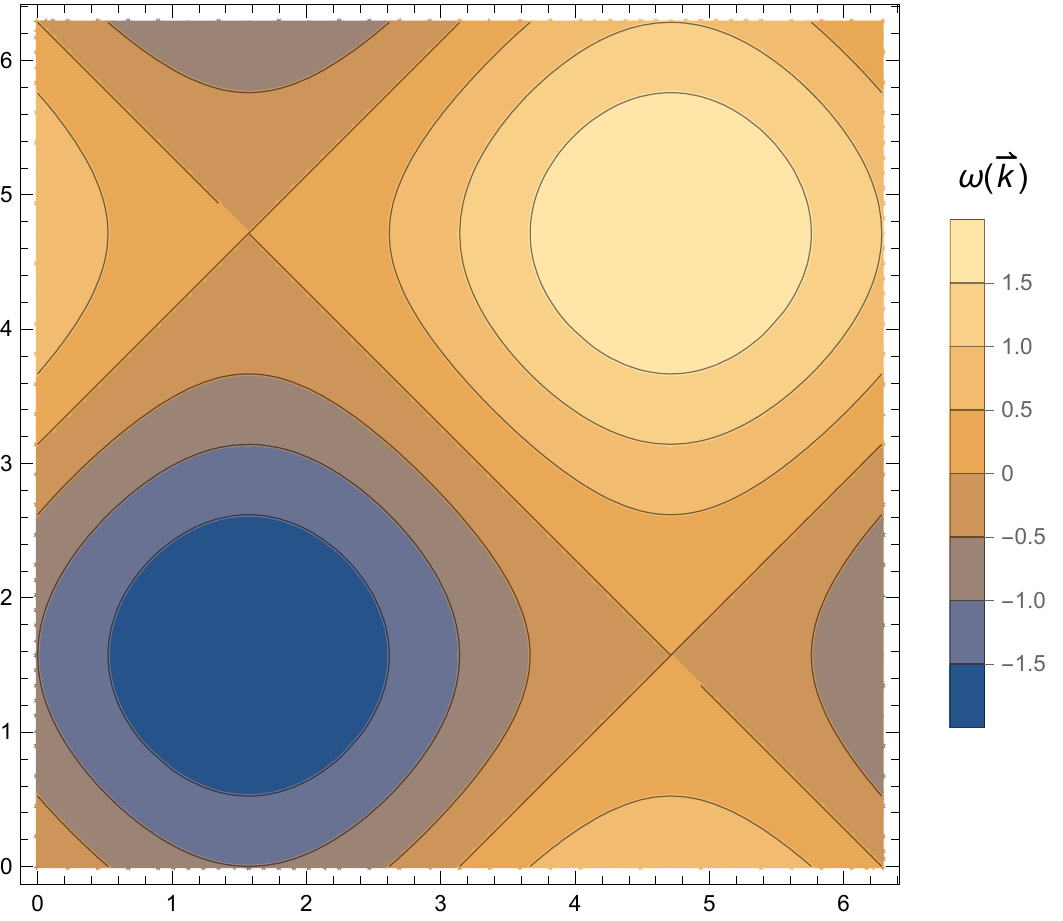}
\caption{Dispersion relation of the half boson model in two dimensions}
\label{fig:disp}
\end{figure}

Let us consider for simplicity a square lattice in two dimensions. We want to develop a model that corresponds to a half boson $q_{i,j}$ per site along the same ideas that we used in one dimension. We want to have translation invariance on the lattice and we want each site to communicate to all its nearest neighbors. Consider the following Poisson bracket
\begin{equation}
\{ q_{i,j}, q_{\ell,m}\} =( \delta_{i,\ell-1} - \delta_{i, \ell+1})\delta_{j,m} + \delta_{i, \ell}(\delta_{j,m-1}-\delta_{j,m+1})\label{eq:PB2d}
\end{equation}
This satisfies what we want: it has the right antisymmetry and has nearest neighbor behavior. Moreover, it essentially copies what we did on rows and columns of the lattice.
If we then consider the Hamiltonian
\begin{equation}
H= \sum_{i,j} \frac 12 q_{i,j}^2
\end{equation}
we can check that it has the following symmetries. It admits a parity reflection symmetry swapping  the $x,y$ directions, basically $P_{swap}: q_{i,j} \to q_{j,i}$. It also admits a parity along the x axis as follows
\begin{equation}
P_X: q_{i,j} \to (-1)^i q_{-i,j}
\end{equation}
combining $P_{swap}$ and $P_X$ one finds that there is also a parity in the $Y$ direction. Together, they also generate a rotation group of the lattice by $90^o$. 
Therefore the system has a hidden rotation invariance, even though the Poisson bracket itself does not make it obvious. Computing the equations of motion is straightforward.
What is more important is the actual dispersion relation, like in \eqref{eq:nosqroot}. We assume that $q_{\ell,m}= \exp(i k_x \ell+ i k_y m-\omega t)$ and find that  
\begin{equation}
\omega(k_1, k_2)= -2 \sin(k_x) -2 \sin(k_y)
\end{equation}
The dispersion relation is shown in figure \ref{fig:disp}.
The reader should notice that the set of zero modes (the locus where $\omega(\vec k)$ vanishes) is of codimension one. This arises because $\omega$ is a {\em single valued function of $k$} and the modes with opposite momentum are canonical conjugates of each other (this is required by translation invariance).
As such if $\omega(k)$ is positive for some $k$, then it is negative for $(-k)$. For any path that connects them, there must be a crossing of zero. This requires that the set of zeros is in codimension one: the boundary of the region positive $w(k)$, which is a level set and cannot be just a critical point.

In this case, it consists of two lines in the torus crossing each other. The first line goes through zero and is antidiagonal. The other one has slope one and crosses through $(\pi,0)$. The most interesting feature is the set of crossing points at $(3\pi/2, \pi/2)$ and
$(\pi/2, 3\pi/2)$ where the two lines intersect. We can expand around $\vec k_0= (3\pi/2, \pi/2)$ as follows:
\begin{equation}
\omega(\vec k_0 +\delta \vec k) \simeq 2\cos(\delta k_x) - 2 \cos(\delta k_y) \sim -\delta k_x^2+ \delta k_y^2
\end{equation}
We can equally well write this as 
\begin{equation}
\omega(\vec k_0 +\delta \vec k) \equiv \delta k_+ \delta k_-
\end{equation}
so that we see that the dispersion relation is a polynomial of the $\delta k$ variables that vanishes on the two lines $ \delta k_+=0$ or $\delta k_-=0$. 

This is similar to the continuous field theory derived in \cite{FIsherBalents} (see also \cite{Seiberg:2020bhn}), which are written in Lagrangian form and lead to a non-trivial dispersion relation with zeros on lines..
 The vanishing of the dispersion relation should be associated to non-trivial symmetries that have non trivial space dependence.

 A similar conclusion follows from the other crossing point. The degrees of freedom at the crossing are the lightest ones apart from the zero modes, as they become quadratic in $\delta k$. Around any other expansion point, the dispersion relation moving away from the zero locus would only be linear in momentum and such modes would be heavier. After all, each factor of $k$ counts as $1/L$ in the lattice size.
Notice that just like in the 1-D system, the mode $a_{\vec k}$ is conjugate to $a_{-\vec k}$. At the special locus, we find that that the two crossing points share the raising and lowering operators, as $\vec k_0 = -\vec k_0'$. The modes of one crossing transform into the other by the  parity symmetry swapping $x,y$. The crossing points are at fixed points of the other parity symmetries, since $k_{0,x}=\pi-k_{0,x} \mod(2\pi)$.

Now we can ask: is there a way to see that the fracton structure is associated to a symmetry?
The answer is yes. We have to go back to how the map from a boson to a half boson on a lattice of twice the size worked: on one set of lattice sites we had momenta, and on the alternating lattice sites we had linear combinations of positions. We notice that in that case the 1-D lattice was bipartite: there was a clear way to separate even and odd lattice sites. 
We want to do the same here. The best way to see that the lattice is bipartite is to think of it as the face centered square lattice, with bonds on the diagonals. This is basically a rotation of the square lattice by $45^0$. 

The vertices of the square are one set of sites, and the center of the faces is the other one.  Let us place momenta at the integer lattice sites $q_{i,j}= p_{i,j}$. 
On the faces we want to write linear combinations of positions that produce the correct signs from \eqref{eq:PB2d}.
We see that  if we choose
\begin{equation}
q_{k+1/2, j+1/2} = -x_{k,j}- x_{k,j+1} + x_{k+1,j} + x_{k+1, j+1},\label{eq:qhalf}
\end{equation}
then the Poisson structure matches what we need.
What are the symmetries here? We get that the $q$ are difference functions along the horizontal direction and sums on the vertical direction. Basically, we can translate all the  $x_{k,j}$ with $k$ fixed by a set amount and none the $q$  change.  We can do the same with the $x_{k,j}$ with $j$ fixed,  but because of sign issues,
the corresponding symmetry on the vertical direction requires that the shifts are Neel ordered (it would be the equivalent of the $\tilde q$ variables that have the shift symmetry).   
 Basically, we have two types of shift symmetry on the  $x$ that can be located either on the rows or the columns and that can be effected with either  arbitrary $x$ dependence or $y$ dependence on the lattice. These symmetries with local parameters in one  direction are a  characteristic property of many fracton models. Here, the Poisson bracket of the $q$ has done it for us automatically.

We can also write models where the Hilbert space is reduced locally to a Hilbert space with a finite number of qubits per unit volume. The process is to replace $p\to \exp(i\alpha p) \sim Q$
and $q_{k+1/2, j+1/2}\to \exp(i \alpha q_{i+1/2, j+1/2}) \simeq P^{-1}_{k,j}\otimes P_{k,j+1}^{-1} \otimes P_{k+1,j}\otimes P_{k+1,j+1}$. These models are very similar to those introduced in \cite{Vijay:2016phm}. 

Notice that in terms of the $P$ variables there is no hopping between nearest neighbors, but something mode complicated instead. Just like in the 1-D model, the zero modes (or equivalently charges) 
\begin{equation}
\Omega_k = \prod_j Q_{j,k}\quad \tilde \Omega_j = \prod_k Q^{(-1)^k} _{j,k},
\end{equation}
commute with the Hamiltonian: the shift symmetries we had in the $x$ that are generated by linear combinations of the $p$.
These depend on one of the coordinates of the lattice. The corresponding discrete exponentials that would produce the translations in  \eqref{eq:qhalf}  survive in a discrete form. 

\section{Conclusion}

In this paper the idea of a half boson degree of freedom on a lattice system was presented. 
The main idea is that the degrees of freedom of the boson end up distributed on the lattice and not located at a single site.
This is encoded in a non-trivial Poisson bracket between neighbors. This is similar to the idea of staggered fermions, where various components of a fermion field reside at different sites.

Here, I studied the simplest such system in one and two dimensions: the systems were constrained in that all lattice sites could be translated into each
 other and that the Poisson bracket and the Hamiltonian was translation invariant. Under those conditions and assuming that the system is free, it turned 
 out that the dispersion relation was a single valued function of the momentum. There is only either only one raising or lowering operator at each $k$ and they have 
 opposite frequencies. This property forces the system to have crossings of zero in the dispersion relation and leads to rich topological features. In two dimensions it led to fracton phases of matter.
 I also showed that when one considers bosonic degrees of freedom that are periodic, the Hamiltonians one produces automatically give rise  to critical spin chain models.
 Systems that are generalizations of  the Toda chain are found. it is interesting to ask if these are integrable. That problem is currently being looked at by the author.
 
 The half boson models seem to have rich topological features that  mimic some of those present in systems of Majorana fermions.
 The simplest topological features of the model (models)  have been noted but there is a lot more to explore. 
 Projections from regular boson degrees of freedom to half bosons were found and they sometimes miss some zero modes of the half bosons. They can also have additional variables left over that do not belong to the half bosons.
 
There should be interesting generalizations of this idea that might be able to mimic other fermion models with interesting topological features, like Kahler-Dirac fermions, 
which can be put on arbitrary triangulations (see for example \cite{Catterall:2018dns}). 
Similarly, even some lattice sites that have translation invariance, can be such that not all lattice sites can be translated into each other: the fundamental cell might contain more than one site. Models on half bosons on these systems might display a richer topology than those that require all lattice site to be equivalent. This is currently under consideration by the author.

\acknowledgements
I would like to thank many discussions with R. Brower, S. Catterall, P. T. Lloyd, Y. Meurice, S. Vijay.
Research supported in part by the Department of Energy under grant DE-SC0019139.

\end{document}